\documentclass[11pt]{article}
\usepackage{graphicx}
\usepackage{amsmath}
\usepackage{dcolumn}
\usepackage{epsfig}
\usepackage{psfrag}

\newcommand{\BABARPubYear}    {08}

\newcommand{\BABARConfNumber} {017}
\newcommand{\SLACPubNumber} {13324}

\input babarsym
\newcommand\vcd {\ensuremath{V_{\mathrm{cd}}}}

\newcommand\vcb {\ensuremath{V_{\mathrm{cb}}}}
\newcommand\vtd {\ensuremath{V_{\mathrm{td}}}}

\newcommand\vtb {\ensuremath{V_{\mathrm{tb}}}}

\newcommand{\e}      [1]   { {\ensuremath{ \times 10^{ {#1} } }}}

\def\etal     {{\it et al.}}

\def\onpeak {on-peak}

\def\kz{\ensuremath{K^0}}
\def\ks{\ensuremath{K^0_S}}
\def\kl{\ensuremath{K^0_L}}
\def\kstzero{\ensuremath{K^{*0}}}
\def\kst    {\ensuremath{K^{*+}}}

\def\piz {\ensuremath{\pi^0}}

\def\jpsiks{\ensuremath{\jpsi\ks}}
\def\jpsikl{\ensuremath{\jpsi\kl}}
\def\jpsikz{\ensuremath{\jpsi\kz}}
\def\psitwosks{\ensuremath{\psitwos\ks}}
\def\etacks{\ensuremath{\etac\ks}}
\def\chiconeks{\ensuremath{\chicone\ks}}
\def\jpsikstzero{\ensuremath{\jpsi\kstzero}}
\def\JpsiKsCh{\ensuremath{\jpsiks(\pi^+\pi^-)}}
\def\JpsiKszz{\ensuremath{\jpsiks(\pi^0\pi^0)}}

\def\cpmodelist{\ensuremath{\jpsiks,\,\jpsikl,\,\psitwosks,\,\etacks,\,\chiconeks,\,\text{and}\, \jpsikstzero}}

\providecommand{\tbline}{\noalign{\vskip 0.04truecm\hrule\vskip0.04truecm}}
\newcommand\dbline{\noalign{\vskip 0.10truecm\hrule}\noalign{\vskip 2pt}  \noalign{\hrule\vskip 0.10truecm}}

\def\mes   {\ensuremath{m_{ES}}}
\def\deltae{\ensuremath{\Delta E}\xspace}

\def\dz   {\ensuremath{\Delta z}}

\def\bb   {\ensuremath{B \overline{B}}}
\def\ifb   {\ensuremath{\mbox{\,fb}^{-1}}\xspace}

\def\Imlambda{\ensuremath{\mathop{\cal I\mkern -2.0mu\mit m}\lambda}}
\def\abslambda{\ensuremath{|\lambda|}\xspace}
\def\Btag{\ensuremath{B_\mathrm{tag}}\xspace}
\def\Brec{\ensuremath{B_\mathrm{rec}}\xspace}
\def\BCP{\ensuremath{B_{\CP}}\xspace}
\def\Bflav{\ensuremath{B_\mathrm{flav}}\xspace}

\def\Gammad{\ensuremath{\Gamma_d}\xspace}
\def\deltaGammad{\ensuremath{\Delta \Gammad}\xspace}

\def\Abar    {\kern 0.20em\overline{\kern -0.20em A}{}\xspace}
\def\Ab      {\ensuremath{\Abar}\xspace}

\def\lepton   {{\sf Lepton}}
\def\kaon     {{\sf Kaon}}
\def\kaonone  {{\sf Kaon I}}
\def\kaontwo  {{\sf Kaon II}}
\def\kaonpion {{\sf Kaon-Pion}}
\def\pion     {{\sf Pion}}
\def\other    {{\sf Other}}
\def\notag    {{\sf Untagged}}

\def\effectiveeta{0.504 \pm 0.033}
\def\fitr{0.233}\def\statr{0.010}\def\systr{0.005}
\def\measurerperp{\fitr \pm \statr\, \stat \pm \systr\, \syst}

\def\lumi {\ensuremath{425.7\, \ifb}}
\def\nbb {\ensuremath{(465\pm 5)\e{6}\, \bb\ \rm pairs}}
\def\nbbx {\ensuremath{(465\pm 5)\e{6}}}
\def\extralumi{211.7}

\def\fittedS{\ensuremath{0.691 \pm 0.029}}
\def\fittedC{\ensuremath{0.026 \pm 0.020}}
\def\rhoSC{\ensuremath{0.3}}

\def\systS{\ensuremath{0.014}}
\def\systC{\ensuremath{0.016}}

\def\fittedstwob{\ensuremath{0.691 \pm 0.029}}
\def\fittedmodl{\ensuremath{0.974 \pm 0.020}}

\def\fittedSks{\ensuremath{0.666 \pm 0.039}}
\def\fittedSkszz{\ensuremath{0.629 \pm 0.092}}
\def\fittedSpsitwoS{\ensuremath{0.905 \pm 0.101}}
\def\fittedSchic{\ensuremath{0.619 \pm 0.161}}
\def\fittedSetac{\ensuremath{0.930 \pm 0.160}}
\def\fittedSkl{\ensuremath{0.698 \pm 0.062}}
\def\fittedSkst{\ensuremath{0.608 \pm 0.241}}

\def\fittedCks{\ensuremath{0.019 \pm 0.028}}
\def\fittedCkszz{\ensuremath{0.093 \pm 0.063}}
\def\fittedCpsitwoS{\ensuremath{0.092 \pm 0.077}}
\def\fittedCchic{\ensuremath{0.133 \pm 0.109}}
\def\fittedCetac{\ensuremath{0.082 \pm 0.125}}
\def\fittedCkl{\ensuremath{-0.030 \pm 0.050}}
\def\fittedCkst{\ensuremath{0.028 \pm 0.084}}

\def\fittedSjpsikz{\ensuremath{0.670 \pm 0.031}}
\def\fittedCjpsikz{\ensuremath{0.019 \pm 0.023}}

\def\fittedSjpsiks{\ensuremath{0.660 \pm 0.036}}
\def\fittedCjpsiks{\ensuremath{0.029 \pm 0.026}}

\def\measuredc{\ensuremath{\fittedC \stat \pm \systC \syst}}
\def\measureds{\ensuremath{\fittedS \stat \pm \systS \syst}}

\long\def\inst#1{\par\nobreak\kern 4pt\nobreak
    {\it #1}\par\vskip 10pt plus 3pt minus 3pt}

\begin{document}

{\pagestyle{empty}

\begin{flushright}
\babar-CONF-\BABARPubYear/\BABARConfNumber \\
SLAC-PUB-\SLACPubNumber \\
\end{flushright}

\begin{center}
\Large \boldmath Update of Time-Dependent \CP Asymmetry Measurements in $b\to \c\overline{c}s$ Decays
\end{center}
\bigskip

\begin{center}
\large The \babar\ Collaboration\\
\mbox{ }\\
\today
\end{center}
\bigskip \bigskip

\begin{center}
\large \bf Abstract
\end{center}
We present updated measurements of time-dependent \CP\ asymmetries in fully reconstructed
neutral \B\ decays containing a charmonium meson.  The measurements reported here use
a data sample of \nbbx\ $\Upsilon(4S) \to \BB$ decays collected with the \babar\ detector
at the PEP-II \B\ factory.  The time-dependent \CP\ asymmetry parameters measured from
\jpsiks, \jpsikl, \psitwosks, \chiconeks, \etacks, and \jpsikstzero\ decays are
\begin{eqnarray}
C_f &=& \measuredc,\nonumber\\
S_f &=& \measureds.\nonumber
\end{eqnarray}

\vfill
\begin{center}

Submitted to the 34$^{\rm th}$ International Conference on High-Energy Physics, ICHEP 08,\\
29 July---5 August 2008, Philadelphia, Pennsylvania.

\end{center}

\vspace{1.0cm}
\begin{center}
{\em Stanford Linear Accelerator Center, Stanford University,
Stanford, CA 94309} \\ \vspace{0.1cm}\hrule\vspace{0.1cm}
Work supported in part by Department of Energy contract DE-AC02-76SF00515.
\end{center}

\newpage
} 

\begin{center}
\small

The \babar\ Collaboration,
\bigskip

%
B.~Aubert,
M.~Bona,
Y.~Karyotakis,
J.~P.~Lees,
V.~Poireau,
E.~Prencipe,
X.~Prudent,
V.~Tisserand
\inst{Laboratoire de Physique des Particules, IN2P3/CNRS et Universit\'e de Savoie, F-74941 Annecy-Le-Vieux, France }
J.~Garra~Tico,
E.~Grauges
\inst{Universitat de Barcelona, Facultat de Fisica, Departament ECM, E-08028 Barcelona, Spain }
% ...

\end{center}\newpage

\section{INTRODUCTION}\label{sec:intro}

The Standard Model (SM) of electroweak interactions describes
\CP violation as a consequence of an
irreducible phase in the three-family Cabibbo-Kobayashi-Maskawa (CKM)
quark-mixing matrix~\cite{ref:CKM}.
In the CKM framework, neutral \B decays to \CP eigenstates containing a
charmonium and a $K^{(*)0}$ meson through tree-diagram dominated processes
provide a direct measurement of
\stwob~\cite{BCP}, where the angle $\beta$ is defined in terms of the CKM matrix
elements $V_{\mathrm{ij}}$ for quarks $i,j$ as $\arg [-(\vcd^{}\vcb^*) / (\vtd^{}\vtb^*)]$.

We identify (tag) the initial flavor of the reconstructed \B
candidate, \Brec, using information from the other \B meson, \Btag, in the
event.
The decay rate $g_+$ $(g_-)$ for a neutral \B meson decaying to a \CP
eigenstate accompanied by a \Bz (\Bzb) tag can be expressed as
\begin{eqnarray}
g_\pm(\deltat) &=& \frac{e^{{- \left| \deltat \right|}/\tau_{\Bz} }}{4\tau_{\Bz} }
\Bigg\{ (1\mp\Delta\mistag) \Bigg.  \pm  (1-2\mistag)
\times \Big [ -\eta_f S\sin(\deltamd\deltat) -
 \Bigg. C\cos(\deltamd\deltat)  \Big] \Bigg\}\:\:\:
\label{eq:timedist}
\end{eqnarray}
where
\begin{eqnarray}
S &=& -\eta_f \frac{2\Imlambda}{1+\abslambda^2},\nonumber\\
C &=& \frac{1 - \abslambda^2 } {1 + \abslambda^2},\nonumber
\end{eqnarray}
the \CP eigenvalue $\eta_f=+1$ ($-1$) for a \CP even (odd) final state,
$\deltat \equiv t_\mathrm{rec} - t_\mathrm{tag}$ is the difference
between the proper decay times of \Brec and \Btag,
$\tau_{\Bz}$ is the neutral \B lifetime, and \deltamd is the mass difference
of the \B meson mass eigenstates
determined from $\Bz$-$\Bzb$ oscillations~\cite{ref:pdg2006}.
We assume that
the corresponding decay-width difference \deltaGammad is zero.
Here, $\lambda=(q/p)(\Ab/A)$~\cite{ref:lambda},
where $q$ and $p$ are complex constants that relate the \B-meson flavor
eigenstates to the mass eigenstates, and $\Ab/A$ is the ratio of
amplitudes of the decay without mixing of a \Bzb or \Bz to 
the final state under study.
The average mistag probability \mistag describes the effect of incorrect
tags, and $\Delta\mistag$ is the difference between the mistag probabilities
for \Bz and \Bzb\ mesons.
The sine term in Eq.~\ref{eq:timedist} results from the interference
between direct decay and decay after $\Bz-\Bzb$ oscillation. A
non-zero cosine term arises from the interference between decay amplitudes
with different weak and strong phases (direct \CP violation $|\Ab/A|\neq 1$) 
or from \CP
violation in $\Bz-\Bzb$ mixing ($|q/p|\neq 1$).
In the SM, \CP violation in mixing and
direct \CP violation in
$b \to \ccbar s$ decays are both
negligible~\cite{ref:lambda}. Under these assumptions, $\lambda=\eta_f
e^{-2i\beta}$, and $C=0$. Thus, the time-dependent \CP-violating asymmetry is
\begin{eqnarray}
A_{\CP}(\deltat) &\equiv& \frac{g_+(\deltat) - g_-(\deltat)}{g_+(\deltat) +
g_-(\deltat)} \\ \nonumber
&=& -(1-2\mistag)\eta_f S \sin{ (\deltamd \, \deltat )},
\label{eq:asymmetry}
\end{eqnarray}
and $S=\stwob$.  If we relax the assumption that $C=0$, then
$S = \sqrt{1-C^2}\stwob$.

Previous \babar\ measurements have reported time-dependent \CP asymmetries
in terms of the parameters \stwob and \abslambda.  In this paper we
report results in terms of $C_f=\eta_f C$ and $S_f = \eta_f S$ to be consistent with 
other time-dependent \CP asymmetry measurements.
We reconstruct \Bz decays to the final states
\jpsiks, \jpsikl, \psitwosks, \chiconeks, \etacks, and
\jpsikstzero\ ($\kstzero\to \ks \pi^0$)~\cite{ref:chargeconj}.  The $\jpsikl$ final state is
\CP even, and the $\jpsikstzero$ final state is an admixture of \CP even
and \CP odd amplitudes.
Ignoring the angular information in $\jpsikstzero$ results in a dilution
of the measured \CP asymmetry by a factor $1-2R_{\perp}$, where
$R_{\perp}$ is the fraction of the $L$=1 contribution.
In Ref.~\cite{ref:rperp} we have measured $R_{\perp} = \measurerperp$,
which gives an effective $\eta_f = \effectiveeta$ for $f=\jpsikstzero$,
after acceptance corrections.  In addition to measuring 
a combined $S_f$ and $C_f$ for the
\CP modes described above, we measure $S_f$ and $C_f$ for each mode individually,
for the $\jpsiks$ mode where we split this into samples with $\ks\to \pi^+\pi^-$ and 
$\pi^0\pi^0$, and for the channel $\jpsikz$ (combining the $\ks$ and $\kl$ final states).
Since our last published result~\cite{ref:babar_sin2beta}, we have added
$82\times10^6\;\BB$ decays and applied improved event reconstruction algorithms
to the entire dataset.

\section{\boldmath THE DATASET AND \babar\ DETECTOR\label{sec:dataset}}
The results presented in this paper are based on data collected 
with the \babar\ detector at the \pep2\ asymmetric energy \epem\ storage rings~\cite{ref:pepcdr}
 operating at the Stanford Linear Accelerator Center. At \pep2, 
9.0 \gev\ electrons and 3.1 \gev\ positrons collide at a center-of-mass
energy of 10.58 \gev which corresponds to the \FourS\ resonance.  
The asymmetric energies result in a boost from the laboratory to the 
center-of-mass (CM) frame of $\beta\gamma\approx 0.56$.
The dataset analyzed has an integrated luminosity of \lumi\
corresponding to \nbb\ recorded at the \FourS\ resonance (\onpeak). 

The \babar\ detector is described in detail elsewhere~\cite{ref:babar_nim}.
Surrounding the interaction point is a five-layer double-sided
silicon vertex tracker (SVT) which measures the impact parameters of 
charged particle tracks in both the plane transverse to, and along 
the beam direction. A 40-layer drift chamber (DCH) surrounds the SVT 
and provides measurements of the momenta for charged 
particles. 
Charged hadron identification is achieved 
through measurements of particle energy-loss in the tracking system 
and the Cherenkov angle obtained from a detector of internally 
reflected Cherenkov light. A CsI(Tl) electromagnetic calorimeter 
(EMC) provides photon detection, electron identification, and 
$\piz$ reconstruction. 
The aforementioned components are surrounded by a solenoid magnet, that provides
a 1.5 T magnetic field.
Finally, the flux return of 
the magnet is instrumented in order to allow discrimination of muons from pions. 
For the most recent $\extralumi\invfb$ of data, a portion of the 
resistive plate chambers constituting the muon system 
has been replaced by limited streamer tubes~\cite{ref:lsta,ref:lstc}.

We use a right-handed coordinate system with the $z$ axis along the electron beam
direction and the $y$ axis upward, with the origin at the nominal beam interaction
point. Unless otherwise stated, kinematic quantities are
calculated in the laboratory rest frame.
The other reference frame which we commonly use is the CM frame 
of the colliding electrons and positrons.

We use Monte Carlo (MC) simulated events generated using the GEANT4~\cite{ref:geant} and EvtGen~\cite{ref:evtgen}
based \babar\ simulation.

\section{\boldmath RECONSTRUCTION OF \B\ CANDIDATES\label{sec:reconstruction}}

We select two samples of events in order to measure the time-dependent \CP\ asymmetry
parameters $S_f$ and $C_f$.  These are a sample of fully reconstructed \B\ meson decays
to flavor eigenstates (\Bflav) and a sample of signal events used in the extraction
of the \CP\ parameters (\BCP).  The \Bflav\ sample consists of \Bz\ decays to
$D^{(*)-}(\pi^+,\,\rho^+,\,a_1^+)$ and $\jpsi \kstzero$ (where $\kstzero\to K^+\pi^-$) final
states.  We use the \Bflav sample of events to determine
dilution parameters (mistag probabilities). The \BCP sample of events consists of
\Bz decays to $\cpmodelist$.
We assume the interference between the \CP side and the tag side
reconstruction is negligible and therefore the dilution parameters are
assumed to be the same for the \Bflav and \BCP samples. 
We also select a sample of fully reconstructed
charged \B\ meson decays to $\jpsi K^+$, $\psitwos K^+$,
$\chicone K^+$, $\etac K^+$, and $\jpsi \kst$ (where $\kst \to
K^+\piz$) final states to use as a control sample.

The event selection is unchanged
from that described in Ref~\cite{ref:babar_sin2beta}.
Events that pass the selection requirements are refined using
kinematic variables.  The \jpsikl\ mode has the requirement that
the difference \deltae\ between the candidate CM energy and the beam energy
$E^*_\mathrm{beam}$ in the CM satisfies $|\deltae| < 80\mev$.  We require that
the beam-energy substituted mass $\mes=\sqrt{(E^*_\mathrm{beam})^2-(p^*_B)^2}$
is greater than $5.2\gevcc$ for all other categories of events, where $p^*_B$
is the \B\ momentum in the \epem\ CM frame.

We calculate the proper  time difference 
\deltat between the two \B decays from the
measured separation \deltaz between the decay vertices of \Brec and \Btag
along the collision ($z$) axis \cite{ref:babarsin2betaprd}.
The $z$ position of the \Brec vertex is determined from the charged
daughter tracks. The \Btag decay vertex is determined by fitting tracks not
belonging to the \Brec candidate to a common vertex,
and including constraints from the beam spot location and the
\Brec momentum~\cite{ref:babarsin2betaprd}.
Events are accepted if the calculated $\deltat$ uncertainty is less than
$2.5\ps$ and $|\deltat|$ is less than $20\ps$. The fraction of signal
MC events satisfying these requirements is $95\,\%$.

\section{\boldmath \B\ MESON FLAVOR TAGGING\label{sec:tagging}}

A key ingredient in the measurement of time-dependent \CP\ asymmetries
is to determine whether 
at the time of the \Btag decay the \Brec\ was a \Bz\ or a
\Bzb. This `flavor tagging'
is achieved with the analysis of the decay products of the recoiling \B\ meson \Btag.
The overwhelming majority of \B\ mesons decay to a final state that is flavor-specific, i.e.
only accessible from either a \Bz\ or a \Bzb, but not from both. The purpose of the flavor tagging
algorithm is to determine the flavor of \Btag\ with the highest possible efficiency $\epsilon_{\rm tag}$
and lowest possible probability \mistag\ of assigning the wrong flavor to \Btag.  It is not necessary
to fully reconstruct \Btag\ in order to estimate its flavor.
The figure of merit for the performance of the tagging algorithm is the effective tagging efficiency
\begin{equation}
Q = \epsilon_{\rm tag} (1-2\mistag)^2,
\end{equation}
which is related to the statistical uncertainty $\sigma_S$ and $\sigma_C$ in the coefficients $S$ and $C$ through
\begin{equation}
\sigma_{S,C} \propto \frac{1}{\sqrt{Q}}.
\end{equation}
We use a neural network based technique~\cite{ref:babarsin2betaprd,ref:babar_sin2beta} that isolates
primary leptons, kaons and pions from \B\ decays to final states containing $D^*$
mesons, and high momentum charged particles from \B\ decays, to determine the flavor of the \Btag.
The output of this algorithm is divided into seven mutually-exclusive categories.
These are (in order of decreasing signal purity) \lepton, \kaonone, \kaontwo, \kaonpion,
\pion, \other\ and \notag.  The performance of this algorithm is
determined using 
the \Bflav sample.
The \notag\ category of events contain no flavor information, so carry no weight in the
time-dependent analysis, and are not used here.

\section{\boldmath LIKELIHOOD FIT METHOD\label{sec:maximum}}

We determine the composition of our final sample by performing simultaneous
fits to the \mes\ distributions for the full \BCP and \Bflav\ samples, except for the
\jpsikl\ sample for which we fit the \deltae\ distribution.

\begin{figure}[!htb]
\begin{center}
\includegraphics[width=0.65\textwidth]{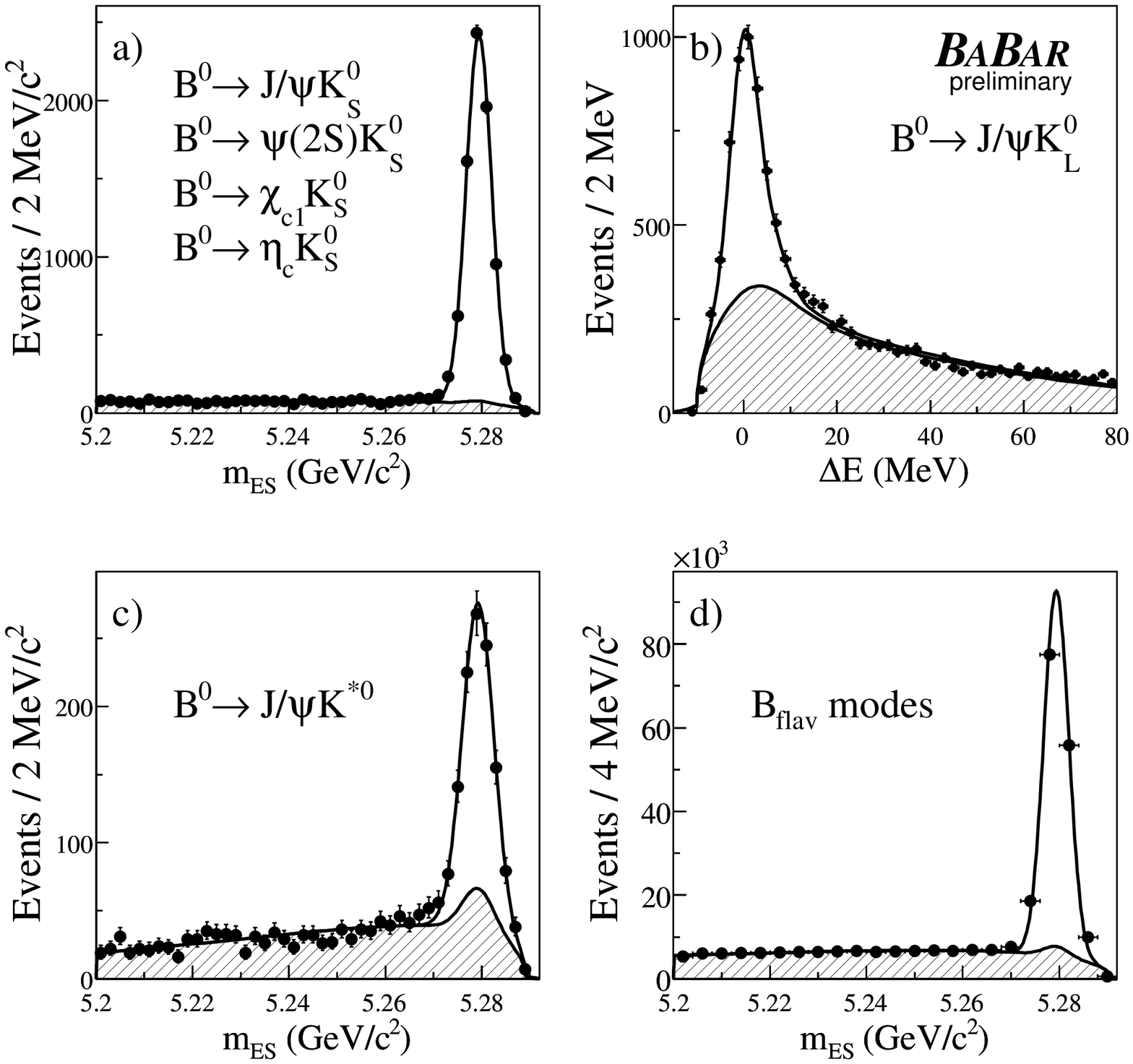}
\caption{
Distributions for \BCP and \Bflav candidates satisfying the tagging and
vertexing requirements:
a) \mes\ for the final states $\jpsi\KS $, $\psitwos\KS$, $\chicone\KS$,
   and $\etac\KS$,
b) \deltae for the final state $\jpsi\KL$,
c) \mes\ for $\jpsi\Kstarz(\Kstarz\to \KS\piz)$, and
d) \mes\ for the \Bflav sample. In each plot, the shaded region is the
   estimated background contribution.
}
\label{fig:bcpsample}
\end{center}
\end{figure}

We define a signal region $5.27 < \mes < 5.29 \gevcc$ ($|\deltae|< 10\mev$ for $\jpsi
\KL$), which contains 15481 \CP candidate events that satisfy the tagging
and vertexing requirements (see Table~\ref{tab:result}).
For all modes except $\etac \ks$ and $\jpsi\kl$, we use simulated events to
estimate the fractions of events that peak in the \mes\ signal region due
to cross-feed from other decay modes (peaking background).
For the $\etac\ks$ mode, the cross-feed fraction is determined
from a fit to the $m_{KK\pi}$ and \mes\ distributions in data.
For the $\jpsi\kl$ decay mode, the sample composition, effective $\eta_f$,
and \deltae\ distribution of the individual background sources are
determined either from simulation (for $B\to\jpsi X$) or from the
$m_{\ellell}$ sidebands in data (for non-\jpsi background).
Figure~\ref{fig:bcpsample} shows the distributions of
\mes\ obtained for the \BCP and \Bflav events, and \deltae\ obtained for 
the \jpsikl\ events.

We determine $S_f$ and $C_f$ from a simultaneous maximum likelihood
fit to the \deltat distribution of the tagged \BCP and \Bflav samples. The
\deltat distributions of the \BCP sample are modeled by
Eq.~\ref{eq:timedist}. Those of the \Bflav sample evolve according to
Eq.~\ref{eq:timedist} with $S_f=C_f=0$.
The observed amplitudes for the \CP asymmetry in the \BCP
sample and for flavor oscillation in the \Bflav\ sample are reduced by the
same factor, $1-2\mistag$, due to flavor mistags.
The \deltat distributions for the signal are convolved with a
resolution function common to both the \Bflav and \BCP samples,
modeled by the sum of three Gaussian functions~\cite{ref:babarsin2betaprd}.
The combinatorial background is incorporated with an empirical description
of its \deltat spectra, containing zero and non-zero lifetime
components convolved with a resolution function~\cite{ref:babarsin2betaprd} distinct
from that of the signal. The peaking background is assigned
the same \deltat distribution as the signal but with $S_f=C_f=0$,
and uses the same \deltat resolution function as the signal.  As the non-zero lifetime
component of the combinatorial background contains both events that are mixed
and un-mixed, we allow the value of $\deltamd$ for this component
to float in the fit.

In addition to $S_f$ and $C_f$, there are 69 free parameters in the
\CP fit. For the signal, these are 
\begin{itemize}
  \item 7 parameters for the \deltat resolution,
  \item 12 parameters for the average mistag fractions \mistag and the differences
        $\Delta\mistag$ between \Bz and \Bzb mistag fractions for each tagging
        category,
  \item 7 parameters for the difference between \Bz and \Bzb reconstruction and
        tagging efficiencies.
\end{itemize}
The background is described by 
\begin{itemize}
 \item 24 mistag fraction parameters,
 \item 3 parameters for the \deltat resolution,
 \item 4 parameters for the  \Bflav time dependence,
 \item 8 parameters for possible \CP violation in the background, 
       including the apparent \CP asymmetry of non-peaking events 
       in each tagging category,
 \item 1 parameter for possible direct \CP violation in
       the $\chicone\KS$ background to $\jpsi\Kstarz$, and 
 \item 3 parameters for possible direct \CP violation in the $\jpsi\KL$ mode, 
       coming from $\jpsi\KS$, $\jpsi\Kstarz$, and the remaining $\jpsi$ backgrounds.
\end{itemize}
The effective \abslambda
of the non-\jpsi background is fixed from a fit to the \jpsi-candidate
sidebands in $\jpsi\KL$. We fix $\tau_{\Bz}=1.530\ps$ and $\Delta m_d = 0.507\ps$~\cite{ref:pdg2006}.
The determination of the mistag fractions and \deltat resolution
function parameters for the signal is dominated by the
\Bflav sample, which is about 10 times more abundant than the \CP sample.

\section{\boldmath LIKELIHOOD FIT VALIDATION\label{sec:fitvalidation}}

Before fitting the data in order to extract \CP\ asymmetry parameters, we
validate the integrity of the likelihood.  We perform three sets of tests
in order to validate the fit.  The first of these tests consists of
generating ensembles of simulated experiments from the PDFs and fitting each
simulated experiment.  The distribution of fitted $S_f$ and $C_f$ parameters
are required to be unbiased, and we verify that the uncertainties are extracted
correctly from the fit by requiring that the distribution of the
pull $\cal P$ on a parameter ${\cal O}$, given by ${\cal P} = ({\cal O}_{\mathrm{fit}} - {\cal O}_{\mathrm{gen}})/\sigma({\cal O}_{\mathrm{fit}})$,
is a Gaussian centered about zero with a width of one.  The quantity ${\cal O}_{\mathrm{fit}}$ is the
fitted value, with a fitted error of $\sigma({\cal O}_{\mathrm{fit}})$, and
${\cal O}_{\mathrm{gen}}$ is the generated value.

The second test involves fitting simulated \CP\ events with the full \babar\ 
detector simulation.
We require that the $\cal P$ distributions for these signal-only simulated experiments
are centered about zero with a width of one.  We assign an systematic uncertainty
corresponding to any deviations and the statistical
uncertainties of the mean values of the fitted  $S_f$ and $C_f$
distributions from the generated values. 

The third test on our ability to extract $S_f$ and $C_f$ correctly is to perform null tests on
control samples of neutral and charged \B events where $S_f$ and $C_f$ should
equal zero.  We use charged
\B decays to $\jpsi K^\pm$, $\psitwos K^\pm$, $\chicone K^\pm$,
$\jpsi K^{*\pm}$ with $K^{*\pm}\to K^\pm \piz$ and $\ks \pi^\pm$,
and neutral \Bflav decays
for this purpose.  The parameters $S_f$ and $C_f$ are zero for these modes
within the SM. 

\section{\boldmath RESULTS\label{sec:results}}

\begin{figure}[!h]
\begin{center}
\includegraphics*[width=0.7\textwidth] {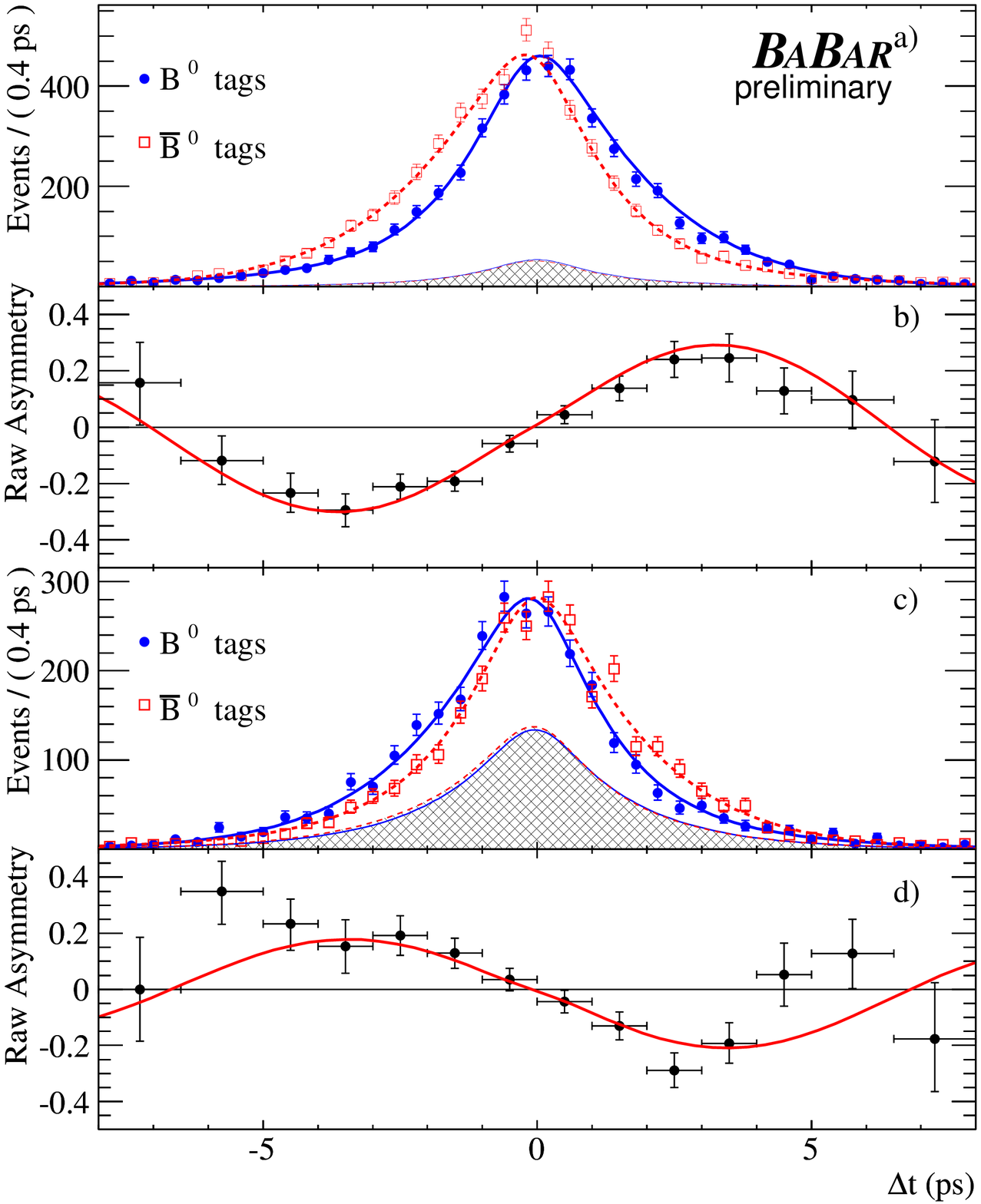}
\caption{
a) Number of $\eta_f=-1$ candidates ($\jpsi\KS$, $\psitwos\KS$,
$\chicone \KS$, and $\etac \KS$) in the signal region with a \Bz tag
($N_{\Bz }$) and with a \Bzb tag ($N_{\Bzb}$), and
b) the raw asymmetry, $(N_{\Bz}-N_{\Bzb})/(N_{\Bz}+N_{\Bzb})$, as functions
of \deltat;
 c) and d) are the corresponding distributions for the $\eta_f=+1$
mode $\jpsi\KL$.
The solid (dashed) curves represent the fit projections in \deltat for \Bz
(\Bzb) tags. The shaded regions represent the estimated background
contributions.}
\label{fig:cpdeltat}
\end{center}
\end{figure}

\begin{figure}[!h]
\begin{center}
\includegraphics*[width=0.7\textwidth] {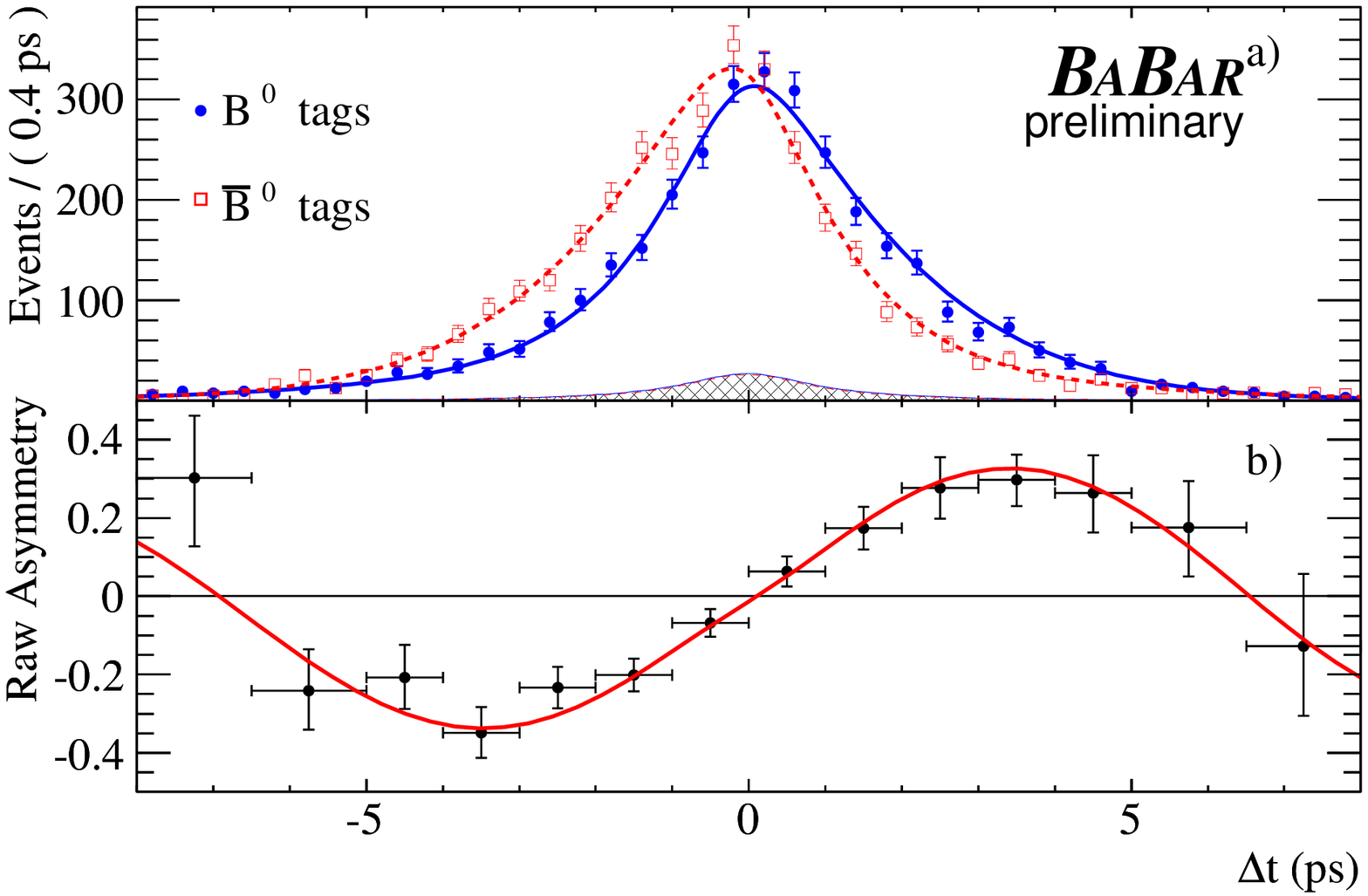}
\caption{
a) Number of $\jpsi\KS$ candidates in the signal region with a \Bz tag
($N_{\Bz }$) and with a \Bzb tag ($N_{\Bzb}$), and
b) the raw asymmetry, $(N_{\Bz}-N_{\Bzb})/(N_{\Bz}+N_{\Bzb})$, as functions
of \deltat.
The solid (dashed) curves represent the fit projections in \deltat for \Bz
(\Bzb) tags. The shaded regions represent the estimated background
contributions.}
\label{fig:cpdeltatjpsiks}
\end{center}
\end{figure}

The fit to the \BCP and \Bflav samples yields $S_f = \fittedS$ and
$C_f = \fittedC$, where the errors are statistical only.
The correlation between these two parameters is $+\rhoSC\,\%$. We also
perform a separate fit in which we allow different $S_f$ and $C_f$
values for each charmonium decay mode, a fit to the
$\jpsi\KS\,(\pipi+\ppz)$ mode, and a fit to the $\jpsi\Kz\,(\KS+\KL)$
sample. We split the data sample by run period and by tagging
category. We perform the \CP measurements on control samples with
no expected \CP asymmetry. The results of these fits are summarized in
Table~\ref{tab:result}.
Figure~\ref{fig:cpdeltat} shows the \deltat distributions and
asymmetries in yields between events with \Bz tags and \Bzb tags for the
$\eta_f=-1$ and $\eta_f = +1$ samples as a function of \deltat,
overlaid with the projection of the likelihood fit result.
Figure~\ref{fig:cpdeltatjpsiks} shows the \deltat distributions and asymmetry
for $\jpsi\ks$ events only.
We also performed the \CP fit using the \stwob and \abslambda parameters, which yields
$\stwob=\fittedstwob$ and $\abslambda=\fittedmodl$.

\begin{table}[!htb]
\vskip-0.4truecm
\begin{center}
\caption{
Number of events $N_{\rm tag}$ and signal purity $P$ in the signal region
after tagging and vertexing requirements, and results of fitting for \CP
asymmetries in the \BCP sample and various subsamples.
In addition, fit results for the \Bflav and $B^+$ control samples demonstrate that
no artificial \CP asymmetry is found where we expect no \CP violation
($S_f=0$, $C_f=0$).
Errors are statistical only.
}
\label{tab:result}
\begin{tabular*}{0.8\textwidth}{@{\extracolsep{\fill}}lrccc}\tbline\tbline
Sample  & $N_{tag}$ & \!$P(\%)$\! & $S_f$      & $C_f$ \\ \tbline
Full \CP sample & 15481 & 76  & \fittedS       & $\phantom{-}\fittedC$ \\ \tbline
\JpsiKsCh       & 5426 & 96   & \fittedSks     & $\phantom{-}\fittedCks$ \\
\JpsiKszz       & 1324 & 87   & \fittedSkszz   & $\phantom{-}\fittedCkszz$ \\
$\psitwos\KS$   & 861 &  87   & \fittedSpsitwoS& $\phantom{-}\fittedCpsitwoS$ \\
$\chicone\KS$   & 385 &  88   & \fittedSchic   & $\phantom{-}\fittedCchic$ \\
$\etac\KS $     & 381 &  79   & \fittedSetac   & $\phantom{-}\fittedCetac$ \\
$\jpsi\KL$      & 5813 & 56   & \fittedSkl     & $           \fittedCkl$ \\
$\jpsi\Kstarz$  & 1291 & 67   & \fittedSkst    & $\phantom{-}\fittedCkst$ \\ \tbline
$\jpsi\Kz$      & 12563 & 77  & \fittedSjpsikz & $\phantom{-}\fittedCjpsikz$ \\ \tbline
$\jpsi\KS$      & 6750 & 95   & \fittedSjpsiks & $\phantom{-}\fittedCjpsiks$ \\ \tbline
$\eta_f=-1$     & 8377 & 93   & $0.688\pm 0.032$ & $\phantom{-}0.041\pm 0.023$ \\ \tbline\hline
1999-2002 data  & 3079 & 78   & $0.736\pm 0.061$ & $\phantom{-}0.013\pm 0.045$  \\
2003-2004 data  & 4916 & 77   & $0.721\pm 0.050$ & $\phantom{-}0.047\pm 0.037$  \\
2005-2006 data  & 4721 & 76   & $0.634\pm 0.051$ & $\phantom{-}0.046\pm 0.035$  \\
2007 data       & 2765 & 75   & $0.666\pm 0.071$ & $-0.017\pm 0.049$ \\ \hline\hline
\lepton         & 1740 & 83   & $0.734\pm 0.052$ & $\phantom{-}0.079\pm 0.038$ \\
\kaonone        & 2187 & 78   & $0.617\pm 0.054$ & $-0.045\pm 0.039$ \\
\kaontwo        & 3630 & 76   & $0.695\pm 0.057$ & $\phantom{-}0.073\pm 0.039$ \\
\kaonpion       & 2882 & 74   & $0.746\pm 0.087$ & $\phantom{-}0.006\pm 0.061$ \\
\pion           & 3053 & 76   & $0.726\pm 0.135$ & $\phantom{-}0.018\pm 0.092$ \\
\other          & 1989 & 74   & $0.767\pm 0.349$ & $-0.168\pm 0.238$ \\ \tbline\tbline
\Bflav sample   & 166276 & 83 & $0.021\pm 0.009$  & $\phantom{-}0.012\pm 0.006$ \\
$B^+$ sample    & 36082  & 94 & $0.021\pm 0.015$  & $\phantom{-}0.013\pm 0.011$  \\\tbline\hline
\end{tabular*}
\end{center}
\end{table}

The dominant systematic errors on $S_f$ are due to limited
knowledge of various background properties, including 
possible differences between the
\Bflav and \BCP tagging performances, 
to the description of the
\deltat resolution functions, 
uncertainties in
$\jpsi\KL$-specific backgrounds and in the amounts of peaking backgrounds
and their \CP asymmetries, 
and to the uncertainties in the values of
the physics parameters $\Delta m_d, \tau_B, \Delta \Gamma_d/\Gamma_d$.
The only sizable systematic uncertainties on $C_f$ are due to 
the \CP content of the peaking backgrounds and to the
possible interference between the suppressed $\bar b\to \bar u c \bar
d$ amplitude with the favored $b\to c \bar u d$ amplitude for some tag-side
\B decays~\cite{ref:dcsd}.
The total systematic error on $S_f$ ($C_f$) is \systS\
(\systC).
The main systematic uncertainties on both $S_f$ and $C_f$ for the full
sample, for the seven individual modes, and for the fits to the $\jpsi\Kz$
and $\jpsi\KS$ samples are summarized in
Tables~\ref{tab:systematicssplitbymodeb} and~\ref{tab:systematicssplitbymode}.

\begin{table*}[!htp]
\vskip-0.4truecm
\caption{Main systematic uncertainties on $S_f$ and $C_f$ for the
full \CP sample, and for the $\jpsi\Kz$, $\jpsi\KS$, and $\jpsi\KL$ samples. For each
source of systematic uncertainty, the first line gives the error on $S_f$
and the second line the error on $C_f$. The total systematic error
(last row) also includes smaller effects not explicitly mentioned in the
table.}
\def\KsCh{\ensuremath{K^{0\scriptstyle (+-)}_{\scriptscriptstyle S}}\xspace}
\def\Ksz{\ensuremath{K^{0\scriptstyle (00)}_{\scriptscriptstyle S}}\xspace}
\def\extParams{\ensuremath{\deltamd, \tau_B, \deltaGammad/\Gammad}\xspace}
\label{tab:systematicssplitbymodeb}
\begin{center}
\begin{tabular*}{0.8\textwidth}{@{\extracolsep{\fill}}lccccc}
\dbline
Source/sample & & Full
       & $\jpsi\Kz$
       & $\jpsi\KS$
       & $\jpsi\KL$ \\ \tbline
                                Beamspot & $S_f$ & 0.0013 & 0.0021 & 0.0027 & 0.0000 \\
                                         & $C_f$ & 0.0006 & 0.0010 & 0.0021 & 0.0001 \\\hline
                      Mistag differences & $S_f$ & 0.0077 & 0.0057 & 0.0059 & 0.0083 \\
                                         & $C_f$ & 0.0047 & 0.0069 & 0.0053 & 0.0052 \\\hline
                      \deltat resolution & $S_f$ & 0.0067 & 0.0068 & 0.0069 & 0.0071 \\
                                         & $C_f$ & 0.0027 & 0.0029 & 0.0034 & 0.0070 \\\hline
                   $\jpsi\KL$ background & $S_f$ & 0.0057 & 0.0063 & 0.0000 & 0.0271 \\
                                         & $C_f$ & 0.0007 & 0.0008 & 0.0000 & 0.0036 \\\hline
     Background fraction                 & $S_f$ & 0.0046 & 0.0034 & 0.0036 & 0.0044 \\
        and \CP content                  & $C_f$ & 0.0029 & 0.0021 & 0.0009 & 0.0107 \\\hline
                  \mes\ parameterization & $S_f$ & 0.0022 & 0.0020 & 0.0026 & 0.0006 \\
                                         & $C_f$ & 0.0004 & 0.0005 & 0.0008 & 0.0002 \\\hline
                              \extParams & $S_f$ & 0.0030 & 0.0033 & 0.0036 & 0.0040 \\
                                         & $C_f$ & 0.0013 & 0.0012 & 0.0011 & 0.0013 \\\hline
                   Tag-side interference & $S_f$ & 0.0014 & 0.0014 & 0.0014 & 0.0014 \\
                                         & $C_f$ & 0.0143 & 0.0143 & 0.0143 & 0.0143 \\\hline
                                Fit bias & $S_f$ & 0.0023 & 0.0044 & 0.0041 & 0.0063 \\
                         (MC statistics) & $C_f$ & 0.0026 & 0.0044 & 0.0041 & 0.0060 \\\hline
\hline                                     	
                                   Total & $S_f$ & 0.0135 & 0.0131 & 0.0119 & 0.0311 \\
                                         & $C_f$ & 0.0164 & 0.0187 & 0.0167 & 0.0270 \\\hline
\hline
\end{tabular*}
\end{center}
\end{table*}

\begin{table*}[!htp]
\vskip-0.4truecm
\caption{Main systematic uncertainties on $S_f$ and $C_f$ 
for the \JpsiKsCh, \JpsiKszz, $\psitwos\KS$, $\chicone\KS$,
$\etac\KS$, and $\jpsi\Kstarz (\Kstarz \to \KS\piz)$ decay
modes. For each
source of systematic uncertainty, the first line gives the error on $S_f$
and the second line the error on $C_f$. The total systematic error
(last row) also includes smaller effects not explicitly mentioned in the
table.}
\def\KsCh{\ensuremath{K^{0\scriptstyle (+-)}_{\scriptscriptstyle S}}\xspace}
\def\Ksz{\ensuremath{K^{0\scriptstyle (00)}_{\scriptscriptstyle S}}\xspace}
\def\extParams{\ensuremath{\deltamd, \tau_B, \deltaGammad/\Gammad}\xspace}
\label{tab:systematicssplitbymode}
\begin{center}
\begin{tabular}{@{\extracolsep{\fill}}lcccccccc}
\dbline
Source/sample & 
       & $\jpsi \KS {(\pipi)}$
       & $\jpsi \KS {(\ppz)}$
       & $\psi(2S) \KS$
       & $\chicone \KS$
       & $\etac\KS$
       & $\jpsi\Kstarz$ \\ \tbline
                                Beamspot &$S_f$ &  0.0027 & 0.0020 & 0.0078 & 0.0284 & 0.0010 & 0.0058 \\ 
                                         &$C_f$ &  0.0017 & 0.0032 & 0.0084 & 0.0115 & 0.0001 & 0.0001 \\ \hline 
                      Mistag differences & $S_f$ & 0.0075 & 0.0074 & 0.0089 & 0.0065 & 0.0064 & 0.0117 \\
                                         & $C_f$ & 0.0039 & 0.0046 & 0.0052 & 0.0067 & 0.0047 & 0.0019 \\ \hline
                      \deltat resolution &$S_f$ &  0.0072 & 0.0074 & 0.0072 & 0.0099 & 0.0163 & 0.0259 \\ 
                                         &$C_f$ &  0.0030 & 0.0043 & 0.0070 & 0.0039 & 0.0036 & 0.0062 \\ \hline 
                   $\jpsi\KL$ background &$S_f$ &  0.0001 & 0.0000 & 0.0001 & 0.0000 & 0.0001 & 0.0001 \\ 
                                         &$C_f$ &  0.0000 & 0.0000 & 0.0000 & 0.0000 & 0.0000 & 0.0000 \\ \hline 
     Background fraction                 &$S_f$ &  0.0032 & 0.0073 & 0.0156 & 0.0174 & 0.0506 & 0.0564 \\ 
                      and \CP content    &$C_f$ &  0.0012 & 0.0034 & 0.0056 & 0.0098 & 0.0187 & 0.0256 \\ \hline 
                  \mes\                  &$S_f$ &  0.0021 & 0.0089 & 0.0238 & 0.0061 & 0.0023 & 0.0372 \\ 
                   parameterization      &$C_f$ &  0.0007 & 0.0063 & 0.0008 & 0.0017 & 0.0005 & 0.0080 \\ \hline 
                              \extParams &$S_f$ &  0.0031 & 0.0073 & 0.0157 & 0.0025 & 0.0158 & 0.0140 \\ 
                                         &$C_f$ &  0.0014 & 0.0013 & 0.0010 & 0.0009 & 0.0020 & 0.0013 \\ \hline 
                   Tag-side interference &$S_f$ &  0.0014 & 0.0014 & 0.0014 & 0.0014 & 0.0014 & 0.0014 \\ 
                                         &$C_f$ &  0.0143 & 0.0143 & 0.0143 & 0.0143 & 0.0143 & 0.0143 \\ \hline 
                                Fit bias &$S_f$ &  0.0048 & 0.0040 & 0.0079 & 0.0072 & 0.0073 & 0.0271 \\ 
                         (MC statistics) &$C_f$ &  0.0042 & 0.0030 & 0.0019 & 0.0042 & 0.0070 & 0.0389 \\ \hline 
 \hline                                   	
                                   Total & $S_f$ & 0.0129 & 0.0179 & 0.0365 & 0.0398 & 0.0566 & 0.0876 \\
                                         & $C_f$ & 0.0160 & 0.0187 & 0.0209 & 0.0257 & 0.0271 & 0.0540 \\ \hline
 \hline
\end{tabular}
\end{center}
\end{table*}

\section{\boldmath CONCLUSIONS\label{sec:conclusions}}
\label{sec:conclusion}

We report improved measurements of the time-dependent \CP\ asymmetry
parameters that supersede our previous results~\cite{ref:babar_sin2beta}.
These measurements are given in terms of $C_f$ and $S_f$ for the first time
with our data sample. We measure
\begin{eqnarray}
C_f &=& \measuredc,\nonumber\\
S_f &=& \measureds,\nonumber
\end{eqnarray}
providing a model independent constraint on the position of the apex of
the Unitarity Triangle~\cite{ref:unitaritytriangle}.  Our measurements
agree with previous published results~\cite{ref:babar_sin2beta,ref:belle_sin2beta}
and with the theoretical estimates of the magnitudes of CKM matrix
elements within the context of the SM~\cite{ref:ciuchini}.  We also report measurements of
$C_f$ and $S_f$ for each of the decay modes within our \CP\ sample and
of the $\jpsi\kz (\ks+\kl)$ sample.

\section{ACKNOWLEDGMENTS}

We are grateful for the 
extraordinary contributions of our \pep2\ colleagues in
achieving the excellent luminosity and machine conditions
that have made this work possible.
The success of this project also relies critically on the 
expertise and dedication of the computing organizations that 
support \babar.
The collaborating institutions wish to thank 
SLAC for its support and the kind hospitality extended to them. 
This work is supported by the
US Department of Energy
and National Science Foundation, the
Natural Sciences and Engineering Research Council (Canada),
the Commissariat \`a l'Energie Atomique and
Institut National de Physique Nucl\'eaire et de Physique des Particules
(France), the
Bundesministerium f\"ur Bildung und Forschung and
Deutsche Forschungsgemeinschaft
(Germany), the
Istituto Nazionale di Fisica Nucleare (Italy),
the Foundation for Fundamental Research on Matter (The Netherlands),
the Research Council of Norway, the
Ministry of Education and Science of the Russian Federation, 
Ministerio de Educaci\'on y Ciencia (Spain), and the
Science and Technology Facilities Council (United Kingdom).
Individuals have received support from 
the Marie-Curie IEF program (European Union) and
the A. P. Sloan Foundation.



\begin{thebibliography}{99}
\bibitem{ref:chargeconj}
Charge conjugation is implied throughout this paper.
\end{thebibliography}
\end{document}